\documentclass[twocolumn,aps,pra,
showpacs,superscriptaddress,nofootinbib,longbibliography]{revtex4-1}
\pdfoutput=1
\usepackage[utf8x]{inputenc}
\usepackage{amsmath}
\usepackage{amsfonts}
\usepackage{amssymb}
\usepackage{makeidx}
\usepackage{cellspace,booktabs}
\usepackage{natbib}
\usepackage{hyperref}
\usepackage{graphicx}
\usepackage{soul}
\usepackage{empheq}

\usepackage{xcolor}
\definecolor{light-gray}{gray}{0.55}

\renewcommand{\dag}{^{\dagger}}
\newcommand{\exv}[1]{ \langle #1 \rangle }

\begin{document}

\begin{abstract}
We report a first-principles study of the driven-dissipative dynamics for Kerr oscillators in the mesoscopic regime. This regime is characterized by large Kerr nonlinearity, realized here using the nonlinear kinetic inductance of a large array of Josephson junctions. The experimentally measured nonlinear resonance lineshapes of the junction array modes show significant deviations from steady-state numerical predictions, and necessitate time-dependent numerical simulations indicative of strong measurement-induced dephasing in the system arising from the large cross-Kerr effect between array modes. Analytical and numerical calculations of switching rate corroborate this by showing the emergence of a slow time scale, that is much longer than the linear decay rate and is set by fluctuation-induced switching times in the bistable regime. Furthermore, our analysis shows that the usual quantum-activated escape treatment is inadequate for prediction of the switching rates at large frequency shifts caused by strong nonlinearities, necessitating a quantum treatment that utilizes the full system Liouvillian. Based on our analysis, we identify a universal crossover parameter that delineates the regimes of validity of semi-classical and quantum descriptions respectively. Our work shows how dynamical switching effects in strongly nonlinear systems provide a unique platform to study quantum-to-classical transitions.
\end{abstract}

\date{\today}
\author{Christian Kraglund Andersen}
\thanks{chanders@phys.ethz.ch}
\affiliation{Department of Physics and Astronomy, Aarhus University, DK-8000 Aarhus C, Denmark}
\affiliation{Department of Physics, ETH Zurich, CH-8093 Zurich, Switzerland}
\affiliation{Institut Quantique and Départment de Physique, Université de Sherbrooke, Québec, Canada J1K 2R1}
\author{Archana Kamal}
\affiliation{Department of Physics and Applied Physics, University of Massachusetts, Lowell, MA 01854, USA}
\affiliation{Research Laboratory of Electronics, Massachusetts Institute of Technology, Cambridge, Massachusetts 02139, USA}
\author{Nicholas A. Masluk}
\thanks{Present address: IBM T.J. Watson Research Center, Yorktown Heights, NY 10598, USA}
\affiliation{Department of Applied Physics, Yale University, New Haven, Connecticut 06511, USA}
\author{Ioan M. Pop}
\thanks{Present address: Physikalisches Institut, Karlsruhe Institute of Technology, 76131 Karlsruhe, Germany}
\affiliation{Department of Applied Physics, Yale University, New Haven, Connecticut 06511, USA}
\author{Alexandre Blais}
\affiliation{Institut Quantique and Départment de Physique, Université de Sherbrooke, Québec, Canada J1K 2R1}
\affiliation{Canadian Institute for Advanced Research, Toronto, Canada}
\author{Michel H. Devoret}
\affiliation{Department of Applied Physics, Yale University, New Haven, Connecticut 06511, USA}

\title{Quantum versus classical switching dynamics of driven-dissipative Kerr resonators}

\maketitle

%ccccccccccccccccccccccccccccccccccccccccccccccccccccccccccccc
%\section{Introduction}
%ccccccccccccccccccccccccccccccccccccccccccccccccccccccccccccc
%
%
Nonlinear optics spans a broad class of phenomena that involve light-induced variation of optical properties of a system. Interestingly, the nonlinearity in optical systems often originates from an inherently quantum mechanical process, but the description of the resulting output can be either classical or quantum, depending on the system under consideration and intensity of the light fields. For instance, nonlinear crystals implementing frequency mixing and stimulated scattering~\cite{boyd2003nonlinear} are described using classical descriptions rooted in average nonlinear material susceptibilities. On the other hand, a new class of effects and non-classical states emerge on quantizing the light fields, which are dealt with in the framework of quantum optics~\cite{carmichael2009statistical}. In order to distinguish between these two descriptions, it is important to understand the transition from quantum to classical dynamics, especially given the burgeoning presence of nonlinear systems in applications triggered by quantum information processing~\cite{nielsen2010quantum, RevModPhys.77.513}.

An example of a nonlinear phenomenon that is of importance in both quantum and classical optics is the Kerr effect, which changes the optical properties in proportion to the intensity of the incident field. Under a coherent drive and single photon loss, the Hamiltonian of the $k$th mode of a physical system subject to the Kerr effect is given by ($\hbar = 1$)
\begin{align}
H_k = \omega_k \, a_k\dag a_k - K_k \, ( a_k\dag a_k )^2 + \varepsilon_k(t)\,a_k\dag + \varepsilon_k^*(t)\,a_k, \label{eq:hamiltonian}
\end{align}
and the dynamics of the mode, under the assumption of Markovian decay, is described by the master equation
\begin{align}
\dot{\rho} = -i [ H_k,\rho] + \kappa_k \Big[a_k \rho a_k\dag - \frac{1}{2} (a_k\dag a_k \rho + \rho a_k\dag a_k) \Big]. \label{eq:mastereq}
\end{align}
Here, $\omega_k$ is the bare resonance frequency mode, $K_k$ is the Kerr coefficient responsible for shifting the mode frequency, $\varepsilon_k(t) = \varepsilon_k e^{-i\omega_d t}$ is the amplitude of a coherent drive with angular frequency $\omega_d$ and $\kappa_k$ is the single-photon decay rate of the mode. We note that the Hamiltonian above assumes that all nonlinear modes are independent; in principle, there can be additional cross-Kerr coupling between different $k$ modes of the system. This driven-dissipative Kerr resonator is ubiquitous in physical systems spanning fiber optics~\cite{agarwal2006fiber}, superconducting quantum circuits with Josephson junctions (JJs)~\cite{Yurke:87,castellanos2008amplification,PhysRevB.86.024503,PhysRevA.86.013814,leib2012networks,Eichler2014,PhysRevApplied.5.024002},  optomechanical systems~\cite{Bose1997,Ludwig2012} and atomic ensembles~\cite{Gupta2007}. 

In particular, JJ-based superconducting circuits naturally realize strong Kerr nonlinearities at the single photon level, which have been extensively utilized both for fundamental quantum optical studies \cite{bozyigit2011antibunching, kirchmair2013observation} and quantum information-inspired applications~\cite{PhysRevA.75.032329, corcoles2015demonstration, kelly2015state}. This breadth of applications is enabled by the highly flexible nature of the nonlinearity realized in superconducting circuits, which can be tuned either in-situ through the flux-tunable SQUID-based designs \cite{Bell2012} or ex-situ through appropriate selection of junction parameters; the latter is usually accomplished by designing junctions with different ratios of Josephson energy and charging energy or, alternatively, by using arrays of junctions~\cite{Eichler2014, PhysRevLett.109.137002, PhysRevB.92.104508}. 

%
%FFFFFFFFFFFFFFFFFFFFFFFFFFFFFFFFFFFFFFFFFFFFFFFFFFFFFFFFFFFFFFFFFFFFFFFFFFFFFFFFFFFFFFFFFFFFFFFFFFFFFFFF
%
\begin{figure}[t!]
\includegraphics[width=\linewidth]{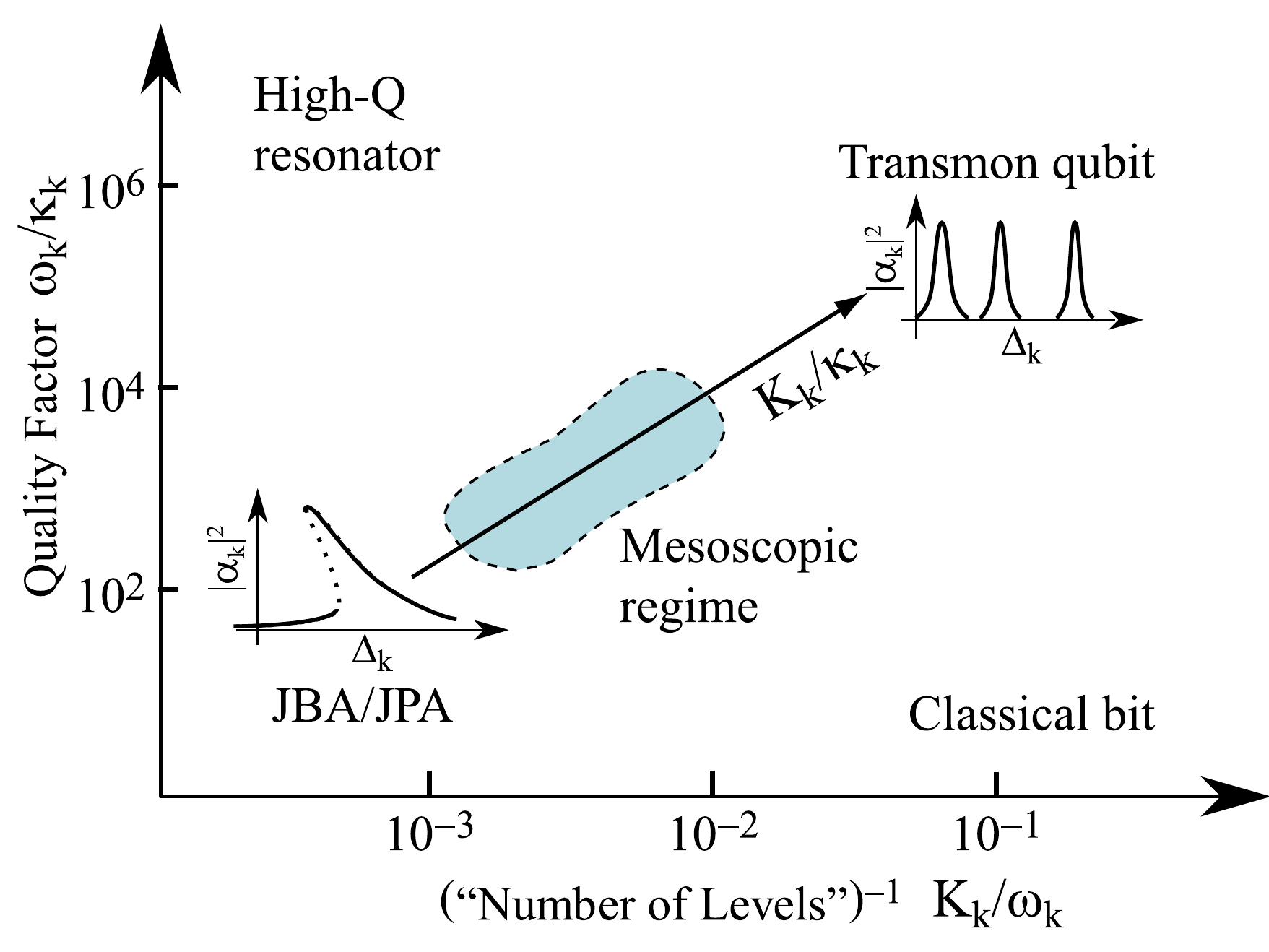}
\caption{Different regimes of Kerr resonator as a function of relative anharmonicity $K_k/\omega_{k}$ which is inversely related to the number of levels in the potential well, and the quality factor $\omega_k / \kappa_k$. The lower left corner corresponds to parametric and bifurcation amplifiers while the upper right corner corresponds to the transmon regime. Each of these regimes is accompanied by a sketch of the average steady-state population of the oscillator, $|\alpha_{k}|^{2}$, in response to a drive amplitude $\varepsilon_{k}$ detuned from resonance by $\Delta_k = \omega_{k} -\omega_{d}$ [see Appendix \ref{app:classical}]. The central region is the intermediate ``mesoscopic" regime that is explored theoretically and experimentally in this work.} \label{fig:kerrregimes}
\end{figure}
%
%FFFFFFFFFFFFFFFFFFFFFFFFFFFFFFFFFFFFFFFFFFFFFFFFFFFFFFFFFFFFFFFFFFFFFFFFFFFFFFFFFFFFFFFFFFFFFFFFFFFFFFFF
%
Figure~\ref{fig:kerrregimes} maps the different nonlinearity regimes readily accessible with JJ circuits. For nonlinearities much smaller than the mode linewidth ($K_k\ll \kappa_k$), the dynamics can be entirely captured by that of a driven classical Duffing oscillator~\cite{dykman2012fluctuating}. This is the domain of the Josephson parametric amplifiers (JPA) \cite{PhysRevLett.109.050507, PhysRevLett.109.050506, lin13singleshot} and Josephson bifurcation amplifiers (JBA) \cite{mallet2009single, vijay2009invited} that are routinely employed for quantum-limited measurements \cite{Murch2013, Sun2014}. Of particular interest in this regime is the bistable behavior occurring when a lossy Kerr resonator is strongly driven. In this bistable regime, the Kerr resonator may switch between two stable states corresponding, respectively, to small and large photon populations of the driven resonator~\cite{muppalla2017bi}. For weak nonlinearity, this switching behavior is well-described by a semi-classical quantum activation theory \cite{vijay2009invited, dykman1980fluctuations, dykman2012fluctuating}. On the other hand, for large nonlinearity ($K_k\gg \kappa_k$), the system enters the photon blockade regime in which the nonlinearity-induced frequency shift limits the number of photons in the oscillator under a drive of fixed frequency. The nonlinear oscillator then becomes an effective two- or `few'-level system; this is the so-called transmon regime \cite{PhysRevA.76.042319}. In contrast to the semi-classical picture, oscillators in this regime do not exhibit any bistability or hysteresis and a full quantum description is necessary to describe their dynamics.

A thorough understanding of fluctuation-induced switching rates in the cross-over region is important for understanding and optimizing Kerr oscillators, both from fundamental physics and application points of view. For instance, nonlinear oscillators exhibit scale invariant behavior near bistability; also, the time needed to switch from one metastable state to the other limits the qubit measurement time in JBAs \cite{mallet2009single, vijay2009invited}. However, the crossover of driven-dissipative dynamics from (semi)-classical to the quantum regime remains poorly understood. In this work, we perform a detailed investigation of this intermediate mesoscopic regime between semi-classical bistability and the transmon regimes for driven Kerr oscillators. The workhorse of our studies is a Kerr oscillator in the strongly nonlinear mesoscopic regime, based on a superinductance formed from an array of Josephson junctions~\cite{VladThesis}. The eigenmodes of the array form highly nonlinear oscillator modes, where the nonlinear shift per photon is larger than the natural oscillator linewidths. Specifically, we use the time scale, here labelled $\tau$, associated with decay into the steady state of the Kerr resonator as a benchmark to delineate semi-classical and quantum dynamics. We report the results of an experiment with a nonlinear resonator realized with a superconducting quantum circuit. The results are simulated using a stochastic master equation and we find signatures of oscillator relaxation time $\tau$ much longer than the intrinsic decay time $1/\kappa$, which motivates further theoretical investigations. We go beyond a linear treatment and present both numerical and analytical calculations of switching rates in this system, considering the situation where the oscillation amplitude is locked in one of the two metastable dynamical states in the presence of a strong drive. Our theoretical studies indicate a breakdown of usual semi-classical treatments that describe oscillator decay primarily as quantum activation in a metapotential or, alternatively, by thermal activation. Instead, in the mesoscopic regime, a quantum treatment is essential to capture the relaxation timsescale of the oscillator.  We characterize this transition from a semi-classical to quantum description by introducing a crossover parameter, $\xi = T_{\gamma}/T_{\kappa}$, as a ratio of a temperature associated with tunneling-induced escape $T_{\gamma}$ and effective temperature associated with fluctuations seen by the oscillator $T_{\kappa}$. When $\xi>1$, quantum effects introduce a new decay channel and fluctuation-induced activation is inadequate to describe the switching dynamics in this regime.

This paper is organized as follows: Section \ref{sec:exp} introduces our experimental implementation of a Kerr resonator in the mesoscopic regime realized by an array of Josephson junctions \cite{PhysRevLett.109.137002}.  In Sec. \ref{sec:theory}, we present the theoretical model describing the ground state properties and Kerr coupling between the distributed modes of the array. By doing a first principle quantization of the array, we confirm that the Kerr coefficients for its distributed modes lie in the regime of interest, $K_k/\kappa_k \approx 2-5$. We then describe how stochastic master equation simulations accurately predict the experimentally observed nonlinear resonance lineshape for an eigenmode of the array in the presence of an external drive. In Secs.~\ref{sec:numr} and \ref{sec:theo}, we perform detailed numerical and analytical investigations of fluctuation-induced switching rates in order to understand the nonlinear effects observed in the experiment. Sec.~\ref{sec:concl} concludes the paper with a discussion of the main results and provides an outlook for future theoretical and experimental studies. Additional numerical and analytical results are included in Appendicies~\ref{app:array} and \ref{app:classical}.

%
%ccccccccccccccccccccccccccccccccccccccccccccccccccccccccccccc
\section{Experiment with Josephson junction arrays}
\label{sec:exp}
\begin{figure}[t]
\begin{center}
\includegraphics[width=\linewidth]{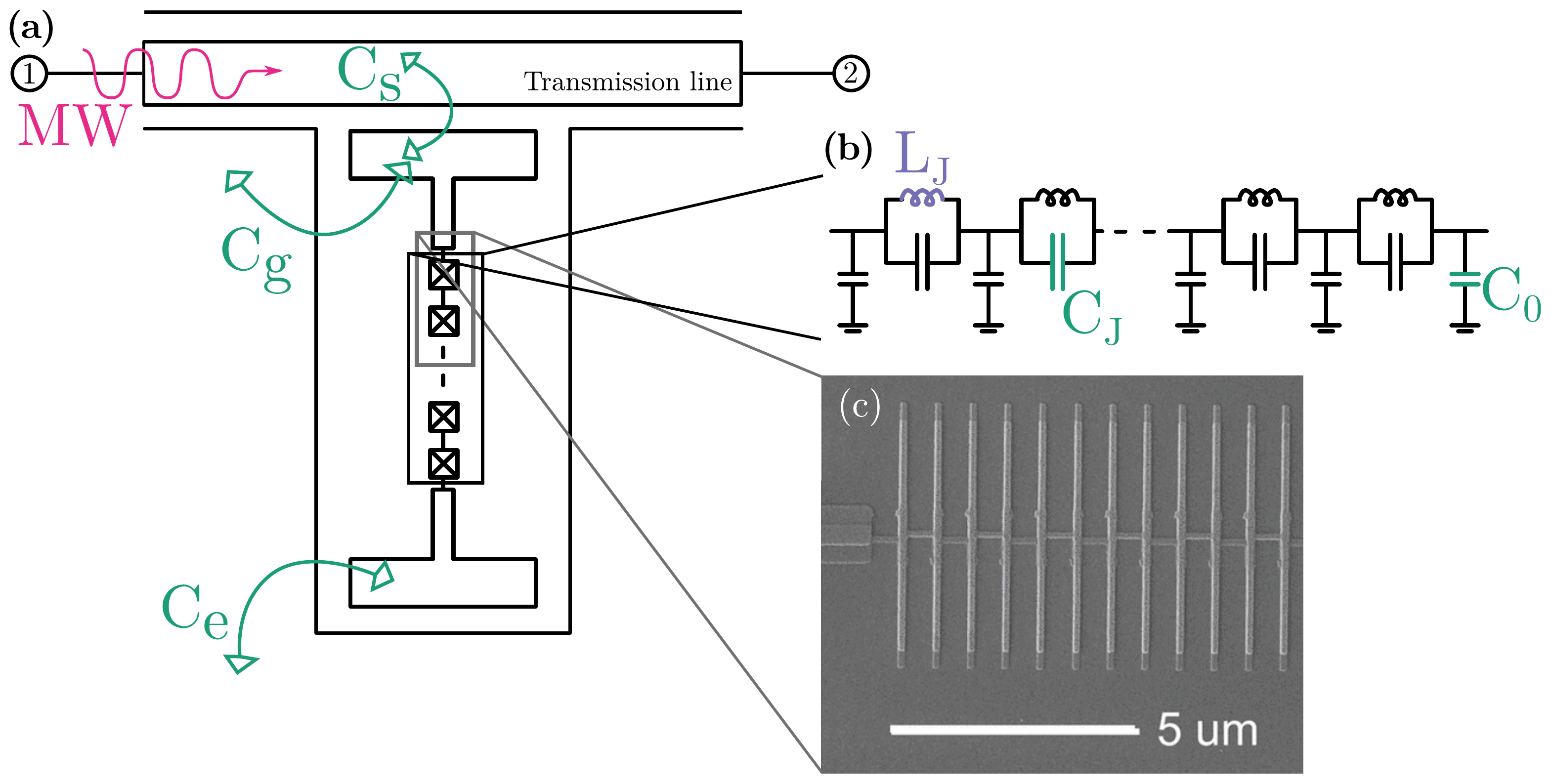}
\caption{(a) Schematic of the experimental set up comprising an array of Josephson junctions coupled to a transmisson line. The capacitance $C_s$ is the coupling capacitance between the array and the tranmission line and $C_g$, $C_e$ represent the capacitances to ground at the two ends of the array. (b) Linear lumped element circuit model for the array. Each junction is represented as an LC-circuit with capacitance $C_J$, inductance $L_J$ and an extra parasitic capacitance to ground $C_0$. (c) Rotated SEM image of the junction array.} \label{fig:experimental_setup}
\end{center}
\end{figure}

%ccccccccccccccccccccccccccccccccccccccccccccccccccccccccccccc
%  
Figure \ref{fig:experimental_setup} depicts the experimental system realizing a mesoscopic Kerr oscillator, which consists of an array of 80 Josephson junctions capacitively coupled to a transmission line in a hanger geometry~\cite{PhysRevLett.109.137002}. The sample is mounted inside a copper box in a dilution fridge at a temperature of 15 mK. Both coherent driving and measurement of the resonator are performed through the transmission line. The transmission signal at the output port~2 is amplified by a HEMT amplifier before being demodulated and recorded. 
\par
%
%FFFFFFFFFFFFFFFFFFFFFFFFFFFFFFFFFFFFFFFFFFFFFFFFFFFFFFFFFFFFFFFFFFFFFFFFFFFFFFFFFFFFFFFFFFFFFFFFFFFFFFFF

%FFFFFFFFFFFFFFFFFFFFFFFFFFFFFFFFFFFFFFFFFFFFFFFFFFFFFFFFFFFFFFFFFFFFFFFFFFFFFFFFFFFFFFFFFFFFFFFFFFFFFFFF
%
The resonator has a fundamental frequency of $\omega_0/2\pi  = 4.357$~GHz, with an internal quality factor of 37,000 and external quality factor of 5,000. Larger internal quality factors were observed for other samples \cite{PhysRevLett.109.137002} but, in all cases, the quality factor was dominated by the external coupling. The experimental setup is constrained by circulators and a low-pass filter to measure signals with frequencies ${<}\,12$ GHz. As a result, only the fundamental mode of the Kerr resonator can be directly probed. Leveraging the nonlinearity-induced mode-mode coupling, it is however possible to probe higher frequency modes. Indeed, a continuous drive applied to the array at the frequencies of its higher modes leads to a shift of the fundamental mode frequency. This shift arises due to the cross-Kerr coupling represented by the interaction term $\sum_{k\neq l} -K_{kl} a_k\dag a_k a_l\dag a_l$ where $K_{kl}$ is the cross-Kerr coupling between array modes $k$ and $l$ while $a_k^{(\dagger)}$ destroys (creates) an excitation in mode $k$. This mode-mode interaction leads to a dispersive readout mechanism for the modes that are above the experimental frequency cutoff (see Appendix~\ref{app:array}). This technique was already used in Ref.~\cite{PhysRevLett.109.137002} to map the full dispersion relation of the distributed modes of the array. In this way, the first mode was identified to be at $\omega_1/2\pi = 11.9$~GHz.
%
%FFFFFFFFFFFFFFFFFFFFFFFFFFFFFFFFFFFFFFFFFFFFFFFFFFFFFFFFFFFFFFFFFFFFFFFFFFFFFFFFFFFFFFFFFFFFFFFFFFFFFFFF

\begin{figure}[t!]
\includegraphics[width=\linewidth]{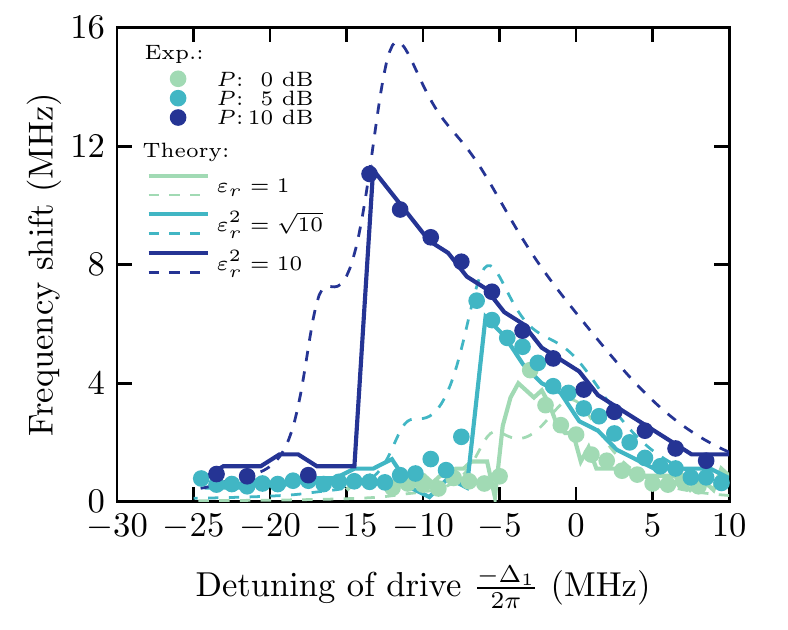}
\caption{Frequency shift of the fundamental mode frequency as a function of pump drive detuning and for different pump drive powers. The dots represent the experimental data while the lines represent the numerical simulations. The legend indicates the power, $P$, of the drive applied both in the experiment and in the simulations, normalized to $0$ dB for the weakest drive display here. For the numerics we have used power proportional to the experimentally applied power, parametrized such that $\varepsilon_r \equiv \varepsilon_1/\varepsilon_{0dB}$ with $\varepsilon_{0dB}/(2\pi) = 1.83$~MHz. The solid lines are a full stochastic two-mode simulation, while the dashed lines are the frequency shifts calculated from an one-mode steady-state assumption using Eqs.~\eqref{eq:mastereq} and~\eqref{eq:Dss}. For the simulation we have used $K_1/(2\pi) = 5.7$~MHz, $\kappa_1/(2\pi) =  2.9$~MHz and $\kappa_0/(2\pi) = 1.0$~MHz. For both experiments and simulations, the frequency shift is obtained by sweeping the probe detuning $\Delta_0$.} \label{fig:data}
\end{figure}

%FFFFFFFFFFFFFFFFFFFFFFFFFFFFFFFFFFFFFFFFFFFFFFFFFFFFFFFFFFFFFFFFFFFFFFFFFFFFFFFFFFFFFFFFFFFFFFFFFFFFFFFF
%
To probe the photon number in this first mode as a function of the drive detuning, a continuous pump tone is applied close to $\omega_1$. Simultaneously, the transmission spectrum of the fundamental mode is measured by sweeping the frequency of a weak continuous probe tone. The frequency at which the probe is maximally reflected indicates the resonance of the fundamental mode.  Figure~\ref{fig:data} shows the experimentally measured frequency shift (dots) obtained using this method for different pump powers and as a function of the detuning of the pump drive. The driven-dissipative Kerr resonator always shows a single resonance peak. Similarly, we note that the measured frequency shifts show abrupt changes between a low and a high value, with no values in between the two. The lack of intermediate values for the frequency shift indicates that no switching occurs between two bistable states during the time of the experiment set by the the measurement time-scale, $t_m = 1 \, \mu$s and averaged 1000 times. We present a detailed theoretical description of the experimentally observed resonance lineshapes in the next section.
%
%%%%%%%%%%%%%%%%%%%%%%%%%%%%%%%%%%%%%
\section{Theoretical model}
\label{sec:theory}
%%%%%%%%%%%%%%%%%%%%%%%%%%%%%%%%%%%%%
%
Since only the fundamental mode is directly measurable experimentally, the Kerr coefficients are inferred from a theoretical model of the distributed junction array. To this end, we write the Lagrangian for the linear part of the array in terms of the system capacitances and Josephson inductance ($L_{J}$), as shown in Fig.~\ref{fig:experimental_setup}(c)~\cite{devoret1995quantum}. Following a quantization of these modes, we perturbatively include the nonlinear contributions from the Josephson junctions (see Appendix~\ref{app:array} for details)~\cite{PhysRevA.86.013814, PhysRevB.92.104508}. From room-temperature resistance measurements, we infer an inductance per junction of $L_J = 1.9$~nH and an effective array plasma frequency (highest mode frequency) of 18.2~GHz. The effective array plasma frequency is not equal to the single junction plasma frequency $\omega_p = 1 / \sqrt{C_J L_J}$, as would be the case if the junctions were purely linear elements. We moreover infer that the system capacitances illustrated in to have values $C_0 = 0.066$ fF, $C_J = 26.54$ fF, $C_g = 10.4$ fF, $C_s = 3$ fF and $C_e = 10.84$ fF close to the design parameters found by simulations~\cite{PhysRevLett.109.137002}. With these parameters, we estimate the self-Kerr coefficient of the fundamental mode to be $K_0/2\pi = 0.5$ MHz, and that for the first mode to be $K_1/2\pi = 5.7$ MHz. The latter is almost a factor of two bigger than the corresponding linewidth $\kappa_1/2\pi = 2.9$ MHz, and hence lies in the intermediate (or mesoscopic) regime identified in Fig. \ref{fig:kerrregimes}. 

Reintroducing the junction nonlinearity in the analysis, we find that the mode frequencies get shifted by self-Kerr and cross-Kerr couplings. The $k$th mode's frequency, $\omega_k$, is shifted such that the new effective frequency becomes
\begin{align}
\omega_k \rightarrow \omega_k - \sum_l K_{kl},
\end{align}
where $K_{kl}$ is the cross-Kerr contribution between modes $k$ and $l$ (see Appendix~\ref{app:array}). Although each $K_{kl}$ is a factor of $10^2$--$10^3$ smaller than $\omega_k$ (on the order of a few MHz), the cumulative effect of all the modes results in significant shifts of the eigenfrequencies of the array by up to a GHz compared to the bare mode frequencies. Moreover, as a result of the cross-Kerr couplings, when driving the Kerr-resonator, the average photon number, $\exv{a_1\dag a_1}$, increases leading to a measurable frequency shift of the fundamental mode. 

As a first, simple, approach to reproducing the data of Fig.~\ref{fig:data}, the steady-state photon number, $\exv{a_1\dag a_1}_{s} = \text{Tr}(\rho_{s} a_1\dag a_1)$, is numerically computed using the single-mode master equation Eq.~\eqref{eq:mastereq}. From this, the frequency shift of the fundamental mode due to photon population in the first mode is then estimated to be
\begin{align}
\Delta_{0} = - K_{01} \exv{a_1\dag a_1}_{s}, \label{eq:Dss}
\end{align}
where $K_{01} = 4\sqrt{K_0 K_1}$ denotes the cross-Kerr coupling between the two modes~\cite{PhysRevA.86.013814}. The result of this calculation corresponds to the dashed lines in Fig.~\ref{fig:data}. Evidently, this steady-state result does not match the experimental data (dots). Indeed only some features of the experiment are captured by this single-mode steady-state calculation. For instance, the detunings are approximately correct. On the other hand, the shape of the frequency shift as a function of the detuning is very poorly reproduced. In particular, the shoulders to the left of the maxima of the shift are not observed experimentally. 

In order to theoretically reproduce the data, we now turn to an approach that resembles the experimental situation more closely. First, rather than computing the steady-state photon population, we compute the full time-dependent response of the transmitted power along the transmission line coupled to the array. As in the experiment, this response is integrated over a finite measurement time $t_m$ and averaged over 1000 realizations. Any dynamics that occur on a time-scale much slower than $t_m$, such as slow switching between two bistable states, are therefore neither resolved in the measurement nor in the simulations. Second, while the cross-Kerr interaction was used in computing the expected frequency shift in Eq.~\eqref{eq:Dss}, we now add this interaction directly to the system Hamiltonian. This allows for capturing the measurement-backaction in the form of measurement-induced dephasing of the first mode by the probe tone. The addition of measurement-induced dephasing increases the effective linewidth of the nonlinear mode, thereby leading to a reduction in the number of photons in the mode and, consequently, to a smaller observed frequency shift.

In order to obtain a description as close to the actual experiment, we numerically solve the full two-mode Hamiltonian including both the mesoscopic Kerr mode ($a_{1}$) and probe mode ($a_{0}$) with the stochastic master equation,
\begin{equation}\label{eq:stochasticmaster}
\begin{split}
d{\rho_Z} &= -i dt\, [H_1 + H_0 + H_c,\rho_Z]\\
&\quad + dt\, \kappa_1 \big[a_1 \rho_Z a_1\dag - \frac{1}{2}(a_1\dag a \rho_Z + \rho_Z a_1\dag a_1) \big] \\
&\quad +  dt\, \kappa_0 \big[a_0 \rho_Z a_0\dag - \frac{1}{2}(a_0\dag a_0 \rho_Z + \rho_Z a_0\dag a_0) \big] \\
&\quad + \sqrt{\kappa_0} \Big[ dZ a_0 \rho_Z + dZ^* \rho_Z a_0\dag\\
&\quad - \text{Tr}(dZ a_0 \rho_Z + dZ^* \rho_Z a_0\dag ) \, \rho_Z \Big]
\end{split}
\end{equation}
valid for heterodyne detection of the fundamental mode $a_0$~\cite{wiseman2009quantum}. In this expression, $H_k$ is Hamiltonian given by Eq.~\eqref{eq:hamiltonian}, while 
\begin{equation}
H_c = -K_{01} \, a_1\dag a_1\, a_0\dag a_0^{\phantom{\dagger}} \label{eq:crosskerrH}
\end{equation}
is the relevant cross-Kerr interaction. In Eq.~\eqref{eq:stochasticmaster}, the density matrix $\rho_Z$ is conditioned on the result of a heterodyne measurement where $dZ = dW_a + idW_b$, with $dW_i$ a stochastic Wiener processes with $\exv{dW_i} = 0$ and $\exv{dW_i^2} = dt$\footnote{The subscripts of $dW_i$ indicates solely that these are different independent Wiener processes.}. 
%The measurement results are included in Eq.~\eqref{eq:stochasticmaster} by the last term. 
The last term of Eq.~\eqref{eq:stochasticmaster} is associated to the homodyne measurement record which is used to update our description of the state of the system $\rho_Z$.
From the numerical integration of Eq.~\eqref{eq:stochasticmaster}, the transmitted power can then be calculated as
\begin{align}
J_T(t) =&\,  \frac{1}{2}\Big\{\sqrt{\kappa_{0}}\,\text{Tr} [\rho_Z (a_0 + a_0\dag)] + dW_a/dt \Big\}^2 \nonumber \\&+ \frac{1}{2}\Big\{i\sqrt{\kappa_{0}}\,\text{Tr} [\rho_Z (a_0 - a_0\dag)] + dW_b/dt \Big\}^2, \label{eq:signalT}
\end{align}
which, averaged over time $t_m$, yields the measured signal. The value $\Delta_0$ corresponding to the maximum transmission signal, $J_T$, gives the frequency shift plotted in Fig.~\ref{fig:data}. The simulation were performed using standard numerical integration for the deterministic part of the equation and after each integration step, $\delta t$, a random number is drawn from a Gaussian distribution with variance $\delta t$ for each stochastic term $dW_i$ such that the stochastic part of $d\rho_Z$ can be readily calculated.

In Fig.~\ref{fig:data}, the numerical simulations using the parameters of the experiment are presented as solid lines, which are in excellent agreement with the experimental data. The difference between the steady-state values used in Eq.~\eqref{eq:Dss} and the stochastic master equation simulations is mainly due to the facts that the latter (i) includes the full two-mode interaction and, as a result, the measurement backaction, and (ii) explicitly takes into account the finite measurement time $t_{m}$. Therefore, dynamics on time-scales much longer than the characteristic measurement time, which are not probed by the experiment, are also not observed in the stochastic master equation simulation. The most predominant feature observed from the stochastic simulations, and not the simpler single-mode steady-state calculations, is the lack of intermediate values in the abrupt transition from low photon number to high photon number state when changing the drive detuning, i.e., the ``shoulders'' in the steady-state photon number. For strong Kerr nonlinearities the energy levels of a Kerr oscillator are {$\omega_k - K_k n$} and can be well-resolved. The ``shoulders'' at large detunings are, thus, signatures of switching events at corresponding photon-number-selective transitions. These intermediate values are not resolved in the experiment due to the finite measurement time and this fact indicates a time-scale of potential switching dynamics much longer than the measurement time $t_m$ and comparably or even longer than the total experimental time of $t_m$ times the number of averages. To fully capture the dynamics of the experiment, a two-mode model is needed, however, the contributions from the two-mode interaction do not significantly contribute to the switching time-scale. Rather, the effects of the cross-Kerr interaction is to account for the measurement-induced backaction. This backaction affects the detuning and drive strength for which the increased time-scales appear. Thus, the physics of the long time-scales can be understood from only a single mode model. In the next two sections, we confirm this by performing a detailed investigation of the switching rates for a single mode Kerr resonator with parameters in the mesoscopic regime, and explicitly comparing the predictions from semi-classical and quantum treatments.
%
%%%%%%%%%%%%%%%%%%%%%%%%%%%%%%%%%%%%%
\section{Numerical treatment of the switching rates}
\label{sec:numr}
%%%%%%%%%%%%%%%%%%%%%%%%%%%%%%%%%%%%%
%
Since both the experiment and stochastic master equation simulations show a qualitatively different behavior than the single-mode steady-state master equation, a more complete quantum description is necessary to quantitatively understand the switching dynamics of Kerr resonator in large-nonlinearity regimes. In particular, as evident from Fig.~\ref{fig:data}, the lack of intermediate photon number values in the bistable regime indicates that there is no dynamical switching between the two bistable states within the time-scale of the experiment. In this section, we perform numerical simulations of the steady-state photon number and switching rates of the Kerr oscillator introduced in the previous section. Specifically, in order to identify the appropriate description for the nonlinear oscillator in mesoscopic regime, we compare the results obtained from semi-classical simulations with that from quantum master equation simulations.

%
%
%FFFFFFFFFFFFFFFFFFFFFFFFFFFFFFFFFFFFFFFFFFFFFFFFFFFFFFFFFFFFFFFFFFFFFFFFFFFFFFFFFFFFFFFFFFFFFFFFFFFFFFFF

\begin{figure}[t]
\includegraphics[width=\linewidth]{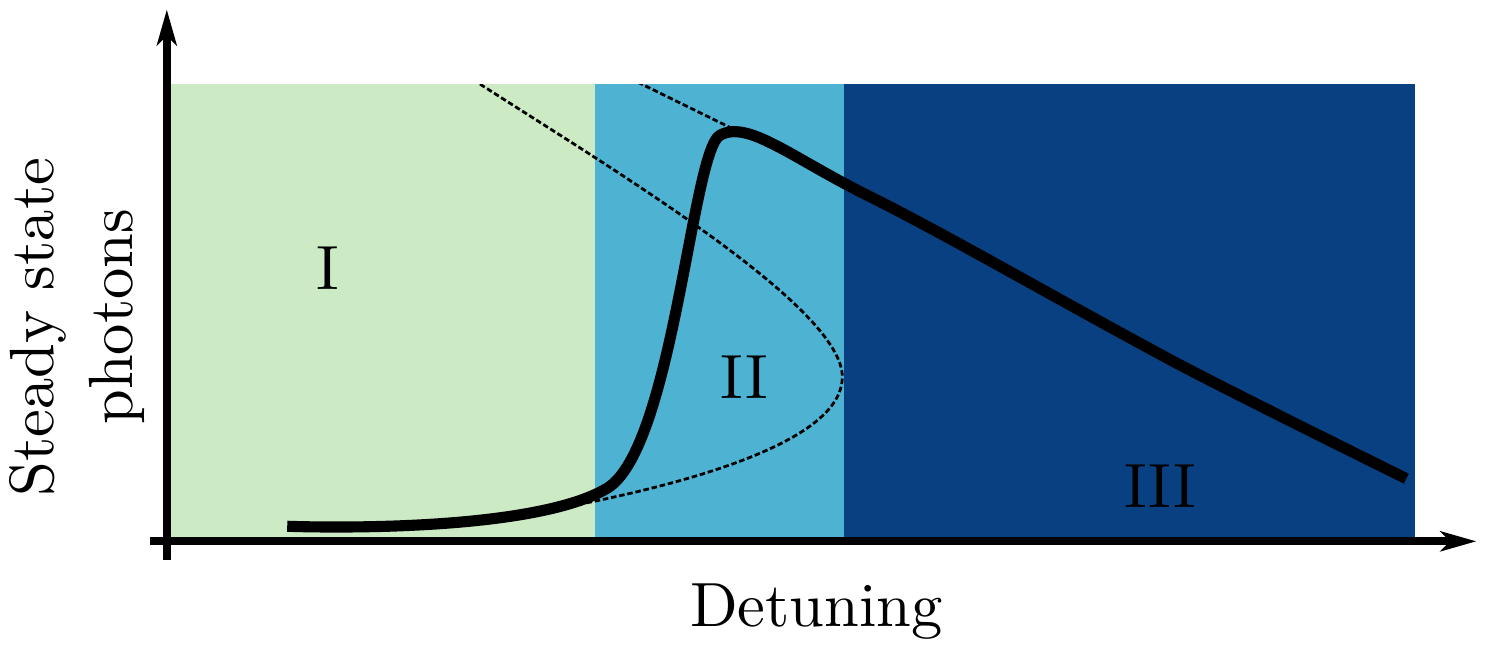}
\caption{Representative curves showing numerical calculations of steady-state photon number as a function of the drive detuning for $K=\kappa$ and $\varepsilon=6\kappa$ (as the red curves in Fig.~\ref{fig:semi_master} but with smaller nonlinearity, i.e., closer to the semi-classical regime). The solid black line shows the results calculated using the master equation, while the dotted lines show the results ontained using a semi-classical equation of motion (see Eq.~\eqref{eq:semi} and Appendix~\ref{app:classical}). In regions I and III, the resonator relaxes to its steady state on an average time scale $T_\kappa = \kappa^{-1}$. Region II corresponds to the switching region where the oscillator dynamics slow down considerably.} \label{fig:regions}
\end{figure}

%FFFFFFFFFFFFFFFFFFFFFFFFFFFFFFFFFFFFFFFFFFFFFFFFFFFFFFFFFFFFFFFFFFFFFFFFFFFFFFFFFFFFFFFFFFFFFFFFFFFFFFFF
%
We begin with a generic description of the steady-state response of a nonlinear oscillator. It is convenient to delineate the time-averaged response into three regions, as sketched in Fig.~\ref{fig:regions} for an oscillator with $K = \kappa$ (this corresponds to the mesoscopic regime depicted as the blue region in Fig.~\ref{fig:kerrregimes}). The thick full line is a representative curve obtained using a master equation simulation, while the narrow dashed line is obtained from a semi-classical analysis with the same parameters. In region I, the resonator relaxes to a state with a low photon number on a time scale of $T_\kappa = 1/\kappa$. There may be two stable classical solutions in this region, as indicated by the dashed line in Fig.~\ref{fig:regions}, but the fluctuations associated with the low photon-number state are not sufficient to bring the system to the high photon number state. For the lower branch solution of this region, the semi-classical and the full quantum treatment give identical results for both the relaxation time into the steady-state and for the steady-state photon number. Indeed, in this regime, only few photons are present and, as a consequence, the nonlinear effects play only a minor role. Region III is conceptually comparable to region I, except that the system relaxes into a high photon number state. 
In region II, the dynamics are such that the photon number initially latches to a low photon-number state, but after some time it jumps to, and continues to fluctuate around, a high photon-number state \cite{Siddiqi2005, PhysRevApplied.1.054005}. This switching between the high and low photon number states continues and we refer to the time-scale of this dynamics as the switching time, $\tau$. We henceforth focus on region II, and study switching times as a function of the ratio $K/\kappa_{1}$. In particular we are interested in quantifying the difference between a weakly nonlinearly Kerr resonator, which we expect to behave classically, and a highly nonlinear Kerr resonator where quantum fluctuations are expected to play a larger role.

We first consider a semi-classical description which we obtain from the Heisenberg-Langevin equation satisfied by the first mode,
\begin{align}\label{eq:hleq}
\dot{a}_{1} = - i \big[a_{1}\, , \, H_{1} \big] - \frac{\kappa_{1}}{2} a_{1} + \sqrt{\kappa_{1}} a_{1}^{\rm in}(t), 
\end{align}
where $H_1$ is given by Eq.~\eqref{eq:hamiltonian}. Here the input field, $a_{1}^{\rm in}(t)$, satisfying the commutation relation $[a_{k}^{\rm in}(t), a_{k'}^{\rm in \dagger}(t') ] = \delta_{kk'}\delta(t - t')$, accounts only for the quantum fluctuations induced by the environment. The driven-dissipative Kerr resonator is most conveniently treated in a frame rotating at the drive frequency, $\omega_d$, such that we make the replacements $\varepsilon_{1}(t) \rightarrow \varepsilon_{1}$ and $\omega_1 \rightarrow \Delta_1 = \omega_1 - \omega_d$ in the Hamiltonian.

We first attempt to solve the time-evolution using a semi-classical trajectory approach \cite{PhysRevA.73.063801},
which replaces the system field operators with complex numbers, $\langle a \rangle \rightarrow \alpha$. This replacements reduces Eq.~\eqref{eq:hleq} to an equation of motion for the phase-space variable $\alpha$ (see also Appendix~\ref{app:classical} for a deterministic classical treatment). However, when replacing the input field with a stochastic variable to account for the quantum noise associated with $a_{1}^{\rm in}(t)$, this approach becomes only approximate in the presence of nonlinear mixing terms in the Hamiltonian, since it does not account for up/down-converted quantum noise due to mode mixing. The resulting semi-classical stochastic equation of motion can then be written as,
\begin{align}
\dot{\alpha} = -i \Delta \alpha +2i\, K |\alpha|^2 \alpha - i \varepsilon - \frac{\kappa}{2} \alpha + \sqrt{\kappa} \, \zeta(t), \label{eq:semi}
\end{align}
corresponding to an equation for the coherent state amplitude of the system, and where $\zeta(t)$ is stochastic Wiener process that models the quantum vacuum noise \footnote{Here, $\zeta(t) = (dW_a(t)/dt + i\, dW_b(t)/dt)/\sqrt{2}$ represents a stochastic Wiener process with $\langle dW_i \rangle = 0$ and $\langle dW_i^2 \rangle = dt$ that, on average, corresponds to the input field being the quantum vacuum state.}. Here, we have suppressed the subscript 1 both for brevity and to underscore the generality of this treatment.
A particular realisation of $\zeta(t)$ is referred to as a semi-classical trajectory. In the bistable regime of the oscillator, a trajectory initially latches to a low photon-number state, and after some time jumps to a high photon-number state \cite{Siddiqi2005, PhysRevApplied.1.054005}. As the switching continues with a waiting time between the successive switches approximately Poisson distributed, we use an exponential function to fit the average over many trajectories to extract the time scale to reach steady-state.

To understand the quantum effects that arise in the switching dynamics of the Kerr resonator, we next perform numerical master simulations, Eq.~\eqref{eq:mastereq}, of the driven system with the Hamiltonian, Eq.~\eqref{eq:hamiltonian}, and analyze the relaxation towards the steady-state. From these simulations, we observe an oscillatory behavior of the photon number that relaxes on a time scale $T_\kappa$ followed by an exponential relaxation towards the steady state. The relaxation time scale, $\tau$, can therefore be readily extracted from the exponent of exponential decay obtained using master equation simulations. 
% 
%FFFFFFFFFFFFFFFFFFFFFFFFFFFFFFFFFFFFFFFFFFFFFFFFFFFFFFFFFFFFFFFFFFFFFFFFFFFFFFFFFFFFFFFFFFFFFFFFFFFFFFFF

\begin{figure}[t]
\includegraphics[width=\linewidth]{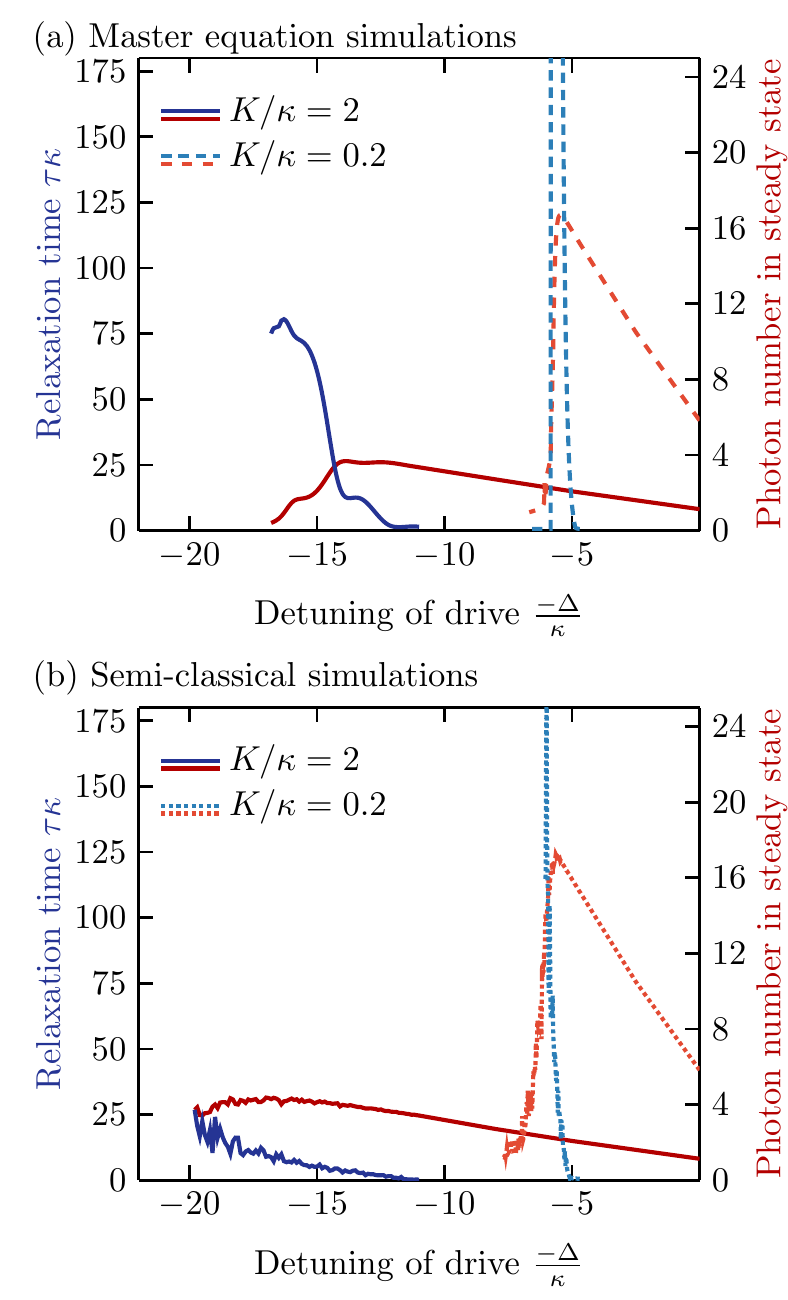}
\caption{Comparison of steady-state photon number and the relaxation time scale $\tau$ (in units of $1/\kappa$) as a function of the drive detuning $\Delta/\kappa$, obtained from quantum master equation simulations (a) and the semi-classical simulations averaged over 200 semi-classical trajectories (b). The red lines represent steady-state photon numbers (right axis), while the blue lines are the time scales obtained from an exponential fit to the time evolution of the photon number (left axis). The solid lines are simulation results for $K=2\kappa$, while the dashed lines show the results for $K=0.2\kappa$. The blue lines indicating the relaxation times $\tau$ associated with the switching dynamics are only plotted in the bistable regime. In all cases, the resonator is initialized in the vacuum state and a drive of amplitude $\varepsilon = 6\kappa$ is used.}
\label{fig:semi_master}
\end{figure}
%FFFFFFFFFFFFFFFFFFFFFFFFFFFFFFFFFFFFFFFFFFFFFFFFFFFFFFFFFFFFFFFFFFFFFFFFFFFFFFFFFFFFFFFFFFFFFFFFFFFFFFFF
%
In Fig.~\ref{fig:semi_master}, we compare the results obtained from both simulations described above, for a strongly nonlinear Kerr oscillator with $K=2\kappa$ (solid lines) and for a weakly nonlinear Kerr oscillator with $K=0.2\kappa$ (dashed lines).  The results of the master equation simulation are shown in the main panel, while those of the semi-classical trajectories are shown in the inset. In both cases, the relaxation time is shown in blue (left axis) and the average photon number in red (right axis). First, it is clear from both the classical and quantum numerical treatments that the switching time (blue lines) can significantly exceed the linear decay time $T_{\kappa}=1/\kappa$ near the bistability. Further, for weak nonlinearity $K=0.2\kappa$, both the quantum and the semi-classical approaches show a sharp rise in $\tau$ close to the detuning where the steady state maximum photon number reaches its maximum (compare dashed lines between the main figure and the inset). This behavior is a generic feature of weakly nonlinear systems. Semi-classically, it can be understood as being caused by the switching rates between two stable states becoming equal. 

For larger nonlinearity, $K=2\kappa$, the discrepancy between the full quantum simulations and semi-classical numerics is much more pronounced (compare solid lines between the main figure and the inset). Specifically, a much wider distribution of time scales is predicted by the quantum treatment in the strong nonlinearity regime, which the semi-classical approach completely fails to capture. This observation is not surprising and is reinforced by the fact that the steady-state in this regime shows a negative $Q$-parameter, \mbox{$Q = (\exv{(\Delta a_1\dag a_1)^2} - \exv{a_1\dag a_1})/\exv{a_1\dag a_1} \approx {-}0.4$}, which indicates sub-Poissonian statistics. Similarly, the steady-state Wigner function displays negative values (not shown), a clear sign of the non-classical nature of the state of the system.  
Numerical simulations with varying $K/\kappa$ further confirm that the average behavior observed in Fig.~\ref{fig:semi_master} gradually transitions from a situation where the semi-classical treatment matches well with full quantum simulation for small $K$, towards a larger discrepancy when $K$ is increased (not shown). 

The numerical results presented here can now be compared to the results of Fig.~\ref{fig:data}. There, the steady-state behavior showed intermediate values in the bistable regime, while the experimental data was only correctly reproduced using a two-mode stochastic quantum analysis that take into account the finite measurement time. This is consistent with the emergence of a very slow time-scale, which can be significantly longer than the linear decay rate $\kappa$ of the oscillator; since dynamics on very long time-scales were not probed by the experiment, it invalidates a steady-state analysis of the data. Moreover, for sufficiently large nonlinearities ($K > \kappa$) as relevant to the experiment with JJ arrays, this slow time-scale cannot be attributed only to semi-classical switching dynamics.
In the next section, we present analytical studies highlighting the differences between the semi-classical picture and the full quantum calculations of the switching rate in the large nonlinearity regime ($K/\kappa \geq 1$), while comparing them against the numerical results obtained in this section. 

%
%%%%%%%%%%%%%%%%%%%%%%%%%%%%%%%%%%%%%
\section{Analytical treatment of the switching rates}
\label{sec:theo}
%%%%%%%%%%%%%%%%%%%%%%%%%%%%%%%%%%%%%
%
%%%%%%%%%%%%%%%%%%%%%%%%%%%%%%%%%%%%%
\subsection{Quantum calculation}
\label{sec:calcquant}
%%%%%%%%%%%%%%%%%%%%%%%%%%%%%%%%%%%%%
%
In the last section, we saw that the increase in the switching time in this regime is not captured by a semi-classical treatment. Therefore, we now consider an analytical description that takes into account the full nonlinear quantum dynamics. 

To this end, we consider the Liouvillian describing the time evolution of the driven-dissipative single-mode Kerr-oscillator in the mesoscopic regime [Eqs.~\eqref{eq:hamiltonian} and~\eqref{eq:mastereq}]. In a Hilbert-space of dimension $N$ ($N$ being the number of Fock states included in the calculation), the Liouvillian has $N^2$ eigenvalues. The time-evolved density matrix can be expressed in terms of these eigenvalues as
\begin{equation}
\rho(t) = \sum_\lambda c_\lambda e^{\lambda t} \rho_\lambda
\end{equation}
where $\rho_\lambda$ are the eigenstates of the Liouvillian, $\mathcal{L}[\rho_\lambda] = \lambda \, \rho_\lambda$. In the presence of damping, all the eigenvalues have negative real parts, with the exception of the steady state $\rho_{s}$  for which $\lambda =0$.  This also represents the only physical state of the system that survives at long times. To preserve the trace at all times, including in steady state, $\rho_{s}$ always appears in this decomposition with a constant prefactor of $c_{s} = 1$. Since the time evolution is trace-preserving this implies that $\text{Tr}(\rho_{\lambda\neq0}) = 0$ and that $\rho_{\lambda\neq0}$ are not valid density matrices with direct physical significance \cite{peres1995quantum}. 

Here, we are interested in the time-dependent fraction of population in the steady state and therefore we re-express the expansion as a linear combination of orthogonal unit-trace density matrices. To accomplish this, we consider the eigenvalue, $\chi \neq 0$ of the Liouvillian, $\mathcal{L}[\cdot]$, defined as the eigenvalue with the smallest real part~\cite{PhysRevA.35.1729}. We observe that $\langle \rho_{s}, \rho_\chi \rangle \neq 0$, with \mbox{$\langle A, B \rangle = \text{Tr}(A\dag B)$} respresenting the matrix inner product. To ensure orthogonality, we use a Gram-Schmidt construction to define
\begin{align}
\tilde{\rho}_\chi = \rho_\chi{-}\frac{\langle \rho_{s}, \rho_\chi\rangle }{\langle \rho_{s}, \rho_{s}\rangle} \rho_{s},
\end{align}
with $\langle \tilde{\rho}_\chi, \rho_{s} \rangle = 0$. The normalized density matrix $\tilde{\rho} = \tilde{\rho}_\chi / \text{Tr}(\tilde{\rho}_\chi)$ has unit trace, but it is not ensured to be positive definite. Nevertheless, we can express the density matrix $\rho(t)$ as
\begin{align}
\rho(t) = \beta_0(t) \rho_{s} + \beta_1(t) \tilde{\rho} + \sigma(t), \label{eq:rhosschi}
\end{align}
where $\sigma (t)$ only ensures the positive definiteness of the full density matrix, while the factors $\beta_{i}$ denote the decay rates. 

\begin{figure}[t!]
\includegraphics[width=\linewidth]{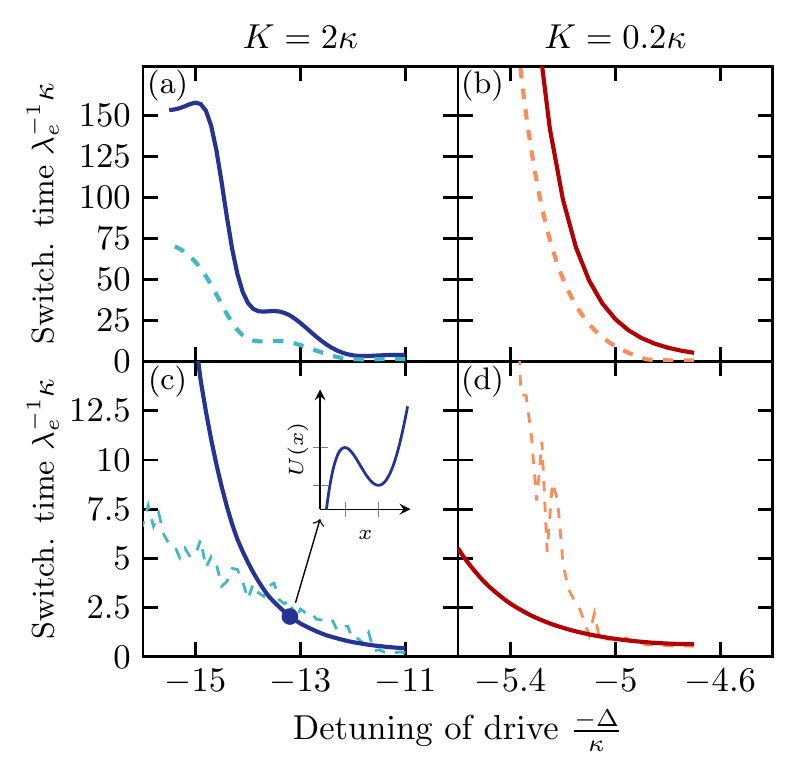}
\caption{Comparison for switching times between oscillators with strong Kerr nonlinearity, $K=2\kappa$, (a) and (c), and weak Kerr nonlinearity $K = 0.2\kappa$, (b) and (d). The panels (a) and (b) present the analytical calculation of the relaxation time scale showed in solid lines, while the dashed lines are the time scales extracted from numerical simulations (also shown in Fig.~\ref{fig:semi_master}). In (c) and (d) the switching time-scales obtained from escape rates in a metapotential are displayed (Eq.~\eqref{eq:lambda_e}]). The dashed lines are the time scales extracted from the semi-classical numerical simulations. Inset: The inset in the lower-left panel displays an example of the metapotential $U(x)$ for the parameters marked by the blue dot. 
The ticks on the x-axis are at 0 and $x_0$ with ticks on the y-axis showing the corresponding values of the energy in the metapotential. 
For all simulations and calculations, driver power $\varepsilon = 6\kappa$ was used.} \label{fig:lambda}
\end{figure}

In order to mimic the exponential relaxation observed in the numerical analysis of Sec. \ref{sec:numr}, we surmise that $\beta_0(t) = (1 - \tilde{\beta}_0 e^{-\lambda_e t})$ with $\lambda_e$ represents the ``escape rate", thereby explicitly assuming Markovian switching dynamics. To estimate the rate at which the density matrix approaches the steady state in the long-time limit, we can calculate the time scale $\lambda_{e}$ by inserting Eq.~\eqref{eq:rhosschi} into the master equation, Eq.~\eqref{eq:mastereq}, and taking the inner product with $\rho_{s}$ to obtain,
\begin{align}
\dot{\beta}_0(t) = \langle \,\rho_{s}, \mathcal{L}[\rho(t)]\, \rangle. 
\end{align}
The quantum-induced switching rate can now be expressed as $\lambda_e = \dot{\beta}_0(t)/(1-\beta_0(t))$. Since $\sigma(t)$ only plays a role in preserving positive definiteness, the dominant time scale is not affected by it and we can omit $\sigma(t)$ in the analysis of the time scale without introducing unphysical behavior. Without $\sigma(t)$, the expression for $\lambda_{e}$ only accounts for the dynamics induced by the steady state and the smallest eigenvalue of $\mathcal{L}[\rho(t)]$, the two dominant contributions in the long time limit. While $\lambda_e$ is explicitly time-dependent, in the regime of interest it is approximately constant and we can therefore neglect the time dependence and simply evaluate for $t=0$ where $\rho(t=0)$ is the initial vacuum state. This leads to the following expression for $\lambda_e$:
\begin{align}
\lambda_e = {\langle \,\rho_{s}, \mathcal{L}[\rho(t=0)]\, \rangle} \Bigg({1 - \frac{\langle \rho_{s}, \rho(t=0) \rangle}{\langle \rho_{s}, \rho_s \rangle}}\Bigg)^{-1}. \label{eq:lambda}
\end{align}
The solid lines in Fig.~\ref{fig:lambda} (a) and (b) show the results of the analytical calculation of the switching times $\lambda_e^{-1}$ for $K/\kappa = 2$ and $K/\kappa = 0.2$, respectively. The dashed lines represent the time-scales extracted from the numerical master equation simulation also shown in Fig.~\ref{fig:semi_master}. Though Eq.~\eqref{eq:lambda} predicts a larger value of $\tau = \lambda_e^{-1}$ than that observed numerically, the analytical calculations qualitatively match the numerical results. We can understand the larger value for $\tau$ obtained analytically from the fact that we neglected $\sigma(t)$. Indeed, in general $\sigma(t)$ has contributions from all eigenvalues of the Liouvillian which includes contributions with a larger negative real part than $\chi$. Therefore, the dynamics associated with $\sigma(t)$ must be strictly faster than $\tau$.  In particular, the Liouvillian spectrum may show a two-fold degeneracy in the real part of the eigenvalues, which may speed up the dynamics by up to a factor of two. The prolonged time-scales observed in both experiment and in numerical simulation can, therefore, directly be qualified using a simple analytical quantum calculation.

%
%FFFFFFFFFFFFFFFFFFFFFFFFFFFFFFFFFFFFFFFFFFFFFFFFFFFFFFFFFFFFFFFFFFFFFFFFFFFFFFFFFFFFFFFFFFFFFFFFFFFFFFFF

%FFFFFFFFFFFFFFFFFFFFFFFFFFFFFFFFFFFFFFFFFFFFFFFFFFFFFFFFFFFFFFFFFFFFFFFFFFFFFFFFFFFFFFFFFFFFFFFFFFFFFFFF
%
%%%%%%%%%%%%%%%%%%%%%%%%%%%%%%%%%%%%%
\subsection{Semi-classical calculation: Quantum activation}
\label{sec:calcclassical}
%%%%%%%%%%%%%%%%%%%%%%%%%%%%%%%%%%%%%
%
The quantum calculation based on the Liouvillian qualitatively reproduces the dynamics of the full master equation. However, for weak nonlinearity, we expect a semi-classical treatment to be sufficient. Here, we compare the switching rates predicted by the numerical simulations with analytical results for a fluctuation-induced escape from a metapotential \cite{dykman2012fluctuating}. A metapotential is an effective potential that corresponds to the same equation of motion as the semi-classical model. 
In this treatment, the fluctuations associated with the noise $\zeta(t)$ in Eq.~\eqref{eq:semi} are transformed into an effective temperature and the switching rate $\tau$ can then be evaluated as a thermal escape rate from a local minimum of the metapotential \cite{dykman1980fluctuations, vijay2008josephson}. The increased time-scale in region II in this picture corresponds to a higher effective barrier in the metapotential.

To estimate the semi-classical escape rate, we begin by first considering the semi-classical equation, Eq.~\eqref{eq:semi}, with $\zeta(t) = 0$ (See Appendix~\ref{app:classical}), and later reinstate the effect of the noise. The general approach is to rephrase the complex equation into a real equation for a generalized position variable, $x$, that changes slowly in time. To this end, we solve Eq.~\eqref{eq:semi} with $\zeta(t) = 0$ and denote the low-amplitude solution as $\alpha_0$ and the unstable solution as $\alpha_u$. Next, we define a rotation angle \mbox{$\varphi = \text{tan}^{-1}[ -\text{Im} (\alpha_u - \alpha_0) / \text{Re} (\alpha_u - \alpha_0)]$} such that the quantity $x_0 = e^{i\varphi} (\alpha_u - \alpha_0)$ is a real number and the axis on the line from 0 to $x_0$ constitutes our position variable $x$. 
In order to determine the metapotential $U(x)$, we make a substitution $\alpha(t)=\alpha_0+e^{-i\varphi}z(t)$ and rewrite the equation of motion, Eq.~\eqref{eq:semi}, in the form $\dot{z}=F(z)$, where
\begin{align}
F(z) =& (-i\Delta - \kappa/2)(z + e^{i\varphi}\alpha_0) - ie^{i\varphi} \varepsilon \nonumber\\&- 2iK (z + e^{-i\varphi}\alpha_0^*)(z + e^{i\varphi}\alpha_0)^2. \label{eq:dUx}
\end{align}
represents the effective force on the particle. The complex variable $z$ can be represented in terms of two real variables, playing the role of the coordinate and momentum, $z=x+ip$.
Since the two states $\alpha_0$ and $\alpha_u$ are steady states, it follows that $F(0) = F(x_0) = 0$. 
Further, since the imaginary part accounts for the dynamics of the momentum $p$, we can write the derivative of the metapotential $U(x)$ as
\begin{align}
\text{Im}[\,F(x)\,] = \dot{p} = - \frac{d U(x)}{d x}.
\end{align}
Integrating this equation results in a one-dimensional metapotential,
\begin{align}
U(x) = - \int \text{Im}[\,F(x)\,] \, dx,
\end{align}
with a minimum at 0 and maximum at $x_0$ as illustrated in the inset of Fig.~\ref{fig:lambda}~(c). Following Kramer's escape law~\cite{kramers1940brownian, RevModPhys.62.251}, the escape rate over the barrier can be written as~\cite{dykman1980fluctuations, vijay2008josephson}
\begin{align}
\lambda_e = \gamma_0 \exp \Big( - \frac{\Delta U}{\kappa} \,\Big), \label{eq:lambda_e}
\end{align}
with \mbox{$\Delta U = U(x_0) - U(0)$} denoting the activation energy. Note that effective temperature is set by the mode linewidth, $\kappa$, since it enters as the prefactor for $\zeta(t)$ in Eq.~\eqref{eq:semi}. The rate $\gamma_0$ is the attempt frequency and it is extracted from the quadratic term of the potential, which can be expressed as $\frac{\gamma_0^2}{2\kappa}x^2$. When increasing the drive amplitude, $\epsilon$, the barrier height is decreased leading to a faster escape rate~\cite{Dykman2007}. 

In Figs.~\ref{fig:lambda} (c) and (d), we show $\lambda_e$ obtained using this quantum-activation approach (solid lines) and compare it with the switching rates extracted from the semi-classical simulations (dashed lines, also shown in the inset of Fig.~\ref{fig:semi_master}). We observe that, for small detunings, the simulations match the escape rate calculation of Eq.~\eqref{eq:lambda_e} quite well. However, a large discrepancy is observed for larger detunings when the switching time dramatically increases near the bifurcation point, where the rates between the two stable states equilibrates as already explained in the context of the numerical simulations. In the escape rate calculation, we are however only calculating a one-way rate and, thus, we do not capture the same increase in switching time. More interestingly, and as should be expected, we observe that while the escape rate calculation captures the behavior of the semi-classical simulations well, it does not capture the quantum corrections relevant in the strongly nonlinear regime ($K/\kappa = 2$), as observed from its deviations from predictions of quantum calculation and master equation simulations [c.f. Figs \ref{fig:lambda} (a) and (c)].

To summarize this section, we have considered the time scales relevant for the relaxation of a Kerr-resonator towards the steady state. Going back to the regions introduced in Fig.~\ref{fig:regions}, we noticed that, most significantly, the time scale for relaxation towards the steady state becomes very large in region II. This is consistent with experimental data presented in Sec.~\ref{sec:exp}, where we found a steady-state description to be inadequate due to the finite time-scale set by the measurement rate (see also the experiments presented in \cite{muppalla2017bi}). We also found that while a semi-classical calculation adequately predicts the dynamics of a weakly nonlinear system, it fails to capture the structure of the time scales for relaxation, $\tau$, seen for large nonlinearities. The time scale for relaxation can be accurately predicted by a simplified but fully quantum model. On the other hand, an escape rate calculation matches  the semi-classical trajectory simulations only partly, and completely fails to describe the dynamics for large nonlinearities. 
%
%%%%%%%%%%%%%%%%%%%%%%%%%%%%%%%%%%%%%
\subsection{Crossover parameter}
\label{sec:crossovertemp}
%%%%%%%%%%%%%%%%%%%%%%%%%%%%%%%%%%%%%
%
The observed breakdown of the semi-classical theory can be understood in a phenomenological manner by introducing a damping-dependent crossover temperature $T_{\gamma}$~\cite{ingold1995},
\begin{eqnarray}
	 T_{\gamma} = \frac{\hbar \gamma_{0}}{2\pi k_{B}} \left(\sqrt{\frac{\kappa^{2}}{4 \gamma_{0}^2}+ 1}-\frac{\kappa}{2 \gamma_{0}}\right),
\end{eqnarray}
where $\gamma_{0}$ is the attempt frequency in the metapotential introduced in the semi-classical calculation. Using this definition, we can define a crossover parameter $\xi$ as
\begin{eqnarray}
	\xi \equiv \frac{T_{\gamma}}{T_{\kappa}},
\end{eqnarray}
where ${T_{\kappa} = \hbar \kappa/k_{B}}$ denotes the effective temperature of the quantum fluctuations coupled to the oscillator. The semi-classical to quantum crossover boundary is set by $\xi =1$~\cite{ingold1995}. 
%
%FFFFFFFFFFFFFFFFFFFFFFFFFFFFFFFFFFFFFFFFFFFFFFFFFFFFFFFFFFFFFFF

\begin{figure}[t!]
\includegraphics[width=0.98\columnwidth]{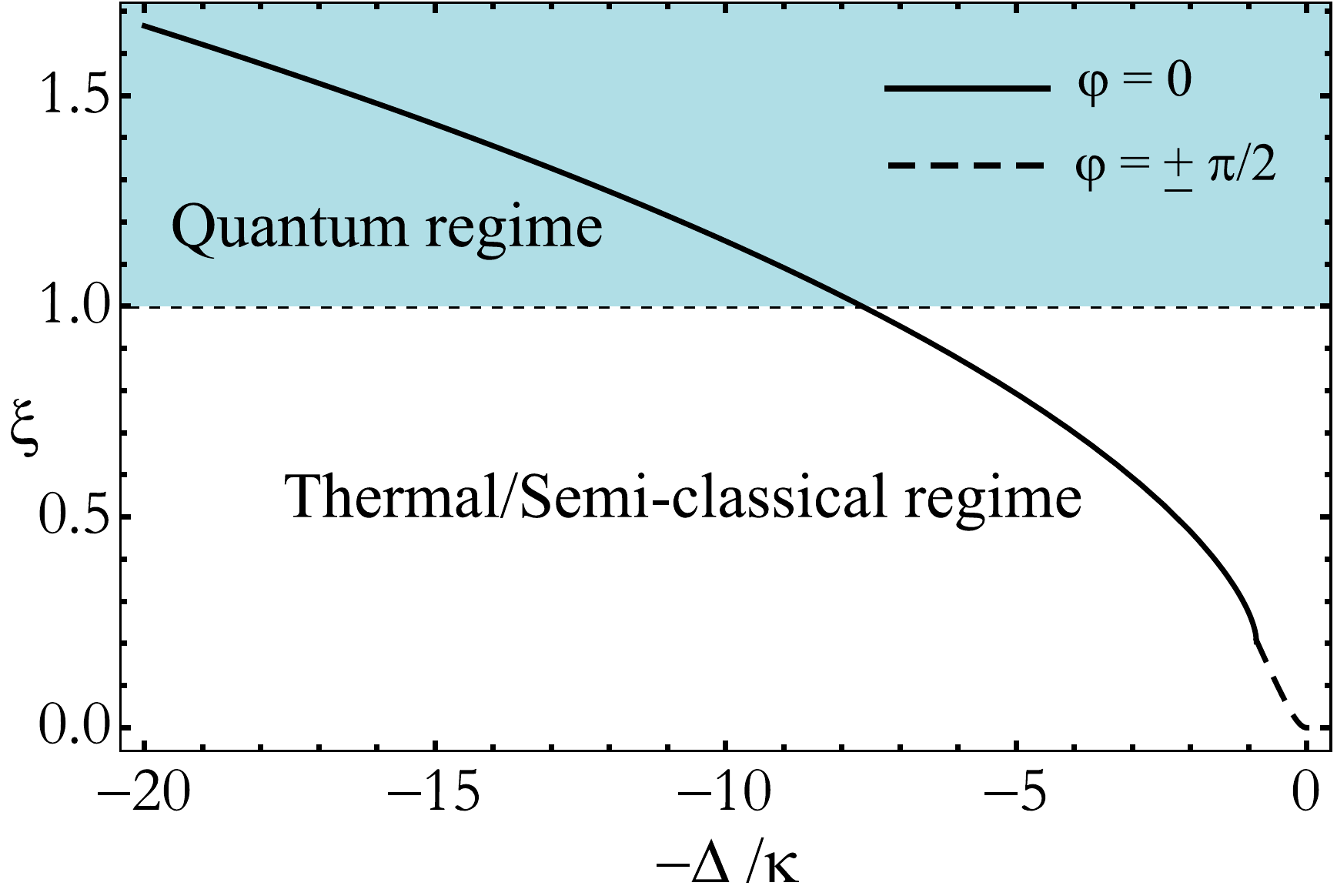}
\caption{Crossover parameter $\xi$ as a function of reduced detuning $\Delta/\kappa$ for a nonlinear oscillator. For $\Delta/\kappa < \sqrt{3}/2$, $\gamma_{0}$ depends entirely on the detuning of the resonator (dashed curve), while for $\Delta/\kappa > \sqrt{3}/2$, $\gamma_{0}$ is largely dominated by the photon amplitude in the resonator (solid curve). This change is captured by the rotation of the real axis of the one-dimensional metapotential denoted by angle $\varphi$. The crossover point is consistent with the region where breakdown of the semi-classical theory was observed previously.} \label{fig:crosstemp}
\end{figure}

%FFFFFFFFFFFFFFFFFFFFFFFFFFFFFFFFFFFFFFFFFFFFFFFFFFFFFFFFFFFFFFFFFFFFFFFFFFFFFFFFFFFFFFFFFFFFFFFFFFFFFFFF
%
Figure \ref{fig:crosstemp} shows $\xi$ as a function of detuning of the oscillator. For $\xi < 1$ or $T_{\kappa} > T_{\gamma}$, the decay dynamics can be described largely using the semi-classical activation treatment. However, as the detuning of the oscillator increases, and ${T_{\gamma} > T_{\kappa}}$, quantum tunneling effects can become essential to describe the nonlinear decay dynamics. We find that the attempt frequency $\gamma_{0}$, and consequently $\xi$, does not depend on the strength of the nonlinearity explicitly. This makes $\xi$ a universal quantity for a dissipative decay of a metastable state; nonetheless, strong nonlinearity $K/\kappa > 1$ is essential for the oscillator to bifurcate at large enough detunings and enter the quantum regime as shown in Fig.~\ref{fig:lambda}. This intuition is consistent with significant deviations of both numerical simulations and Liouvillian-based analytical estimates from quantum activation results at large nonlinearity. Nonlinear oscillators with large $K/\kappa$ thus access a qualitatively different regime from that described by (thermal or quantum) activation-based models, in which rate of escape via tunneling is exponentially small \cite{Dykman2007}. 

%
%ccccccccccccccccccccccccccccccccccccccccccccccccccccccccccccc
\section{Discussion and Outlook}
\label{sec:concl}
%ccccccccccccccccccccccccccccccccccccccccccccccccccccccccccccc
%
In conclusion we have investigated the switching dynamics of a Kerr resonator in a combined experimental, numerical and analytical framework. We used experimental microwave spectroscopy data of a distributed mode of a Josephson junction array with $K/\kappa \approx 2$ and observed that the relaxation into the quantum steady state is not resolved in the experiment due to switching-times much longer than the probe time. 
%We explain this lack of structure by a time-scale argument in that the measurements are dephasing the system much faster than the slow time scale associated with the nonlinear quantum effects. 
We also simulate the experiment using a two-mode stochastic master equation and find that the numerical simulations match very well with the experimental data, thus confirming our interpretation. A recent experiment \cite{muppalla2017bi} has directly measured the very slow switching dynamics in a weakly nonlinear Kerr resonator.

To further analyze the slow switching rates observed in the experiment we performed both semi-classical trajectory simulations and quantum master-equation simulations. We find that, for a range of parameters, the time scale to relax into the steady state is increased significantly beyond the natural decay time of the resonator, especially for strong nonlinearity ($K/\kappa \geqslant 1$) as compared to that for weak nonlinearity  ($K/\kappa \ll 1$). We find that a semi-classical trajectory method is able to describe this slowdown only in the weak nonlinearity regime, while a full quantum master equation treatment is essential to calculate the switching rates for strongly nonlinear oscillators. To analytically estimate the time scales to reach the steady state, we compared a simplified quantum model and a semi-classical metapotential model. In the semi-classical metapotential treatment, the fluctuation-induced switching between bistable states is modelled as an activation over barrier in a metapotential, with the fluctuation-intensity determined by thermal or quantum noise. We find significant deviations from this model, especially when switching rates are small; moreover, these deviations are especially pronounced for large $K /\kappa$ and persist even far from bifurcation. In contrast, we obtain good qualitative agreement between quantum calculations and full master equation simulation in this regime. This is not entirely surprising since thermal or quantum activation models necessarily assume weak nonlinearity or weak driving conditions for the oscillator \cite{Dykman2007}. Our results indicate that switching dynamics in mesoscopic oscillators i.e. when $K/\kappa \gtrsim 1$ may be dominated by some other mechanism, such as dynamical quantum tunneling. 

The conclusions drawn in this work regarding the time scales associated with switching dynamics in nonlinear Kerr-resonators are highly relevant for the characterization of state-of-the-art applications of Josephson devices especially in the strong nonlinearity regimes where quantum effects are more pronounced. In particular, we describe how the interplay of switching rates and the repetition rate of the experiment is essential to explain the experimentally measured nonlinear response of the superinductance presented in Fig. 3. We expect our results to guide the design of applications that similarly aim to observe non-linear spectroscopic signatures in Kerr resonators. In addition, our results can directly be used for assessing the performance of a bifurcation readout scheme for superconducting qubits~\cite{mallet2009single, vijay2009invited}. In such a scheme, the required measurement time has to be a few multiples of the characteristic switching time. Thus, Eqs.~\eqref{eq:lambda} and \eqref{eq:lambda_e} can be directly used for finding the required measurement time and, consequently, the expected qubit readout fidelity. These concerns become increasingly important as devices based on arrays of Josephson junctions, similar to the application in this work, continue to play an important role in experiments and theory proposals including various readout schemes~\cite{castellanos2008amplification, PhysRevA.90.062333, PhysRevLett.113.110502}, quantum controllers \cite{PhysRevA.93.012346, PhysRevX.5.041020, PhysRevApplied.4.034002} and even qubit architectures \cite{Bell2012,PhysRevLett.113.247001,Dempster2014}. 

Furthermore, our calculations for the Kerr coefficients for the higher distributed modes of the Josephson array presented here (see Appendix~\ref{app:array}) indicate that these modes should be in the `mesoscopic' regime (defined as $K \gtrsim \kappa$) investigated here. This regime is also optimal for direct observation of dissipative quantum tunneling~\cite{Serban2007}, which is usually obscured by the activation-dominated switching observed in the JBA regime~\cite{Siddiqi2005}. Moreover, since the switching rate does not follow activation dependence on noise intensity, this suggests that mesoscopic Kerr oscillators can be useful platforms to test dynamics resulting from non-Gaussian noise. More generally, our study provides a framework to explore multi-photon quantum effects \cite{Kryuchkyan2012}, quantum noise properties, and quantum-to-classical transitions in strongly nonlinear systems. 
%ccccccccccccccccccccccccccccccccccccccccccccccccccccccccccccc
\section*{Acknowledgements}
%ccccccccccccccccccccccccccccccccccccccccccccccccccccccccccccc
%
The authors acknowledge useful and insightful discussions with E.~Doucet, S.~Boutin and G.~Kirchmair. CKA acknowledges fruitful discussions with K. Mølmer, the hospitality of Universit\'{e} de Sherbrooke and financial support from the Villum Foundation and the Danish Ministry of Higher Education and Science. The authors acknowledge support by the U.S. Army Research Office under the grant number W911NF-18-1-0212, by the U.S. Department of Energy under grant number DE-SC0019515, by NSERC, and in part by funding from the Canada First Research Excellence Fund. 
% \hl{Please write everything that needs to be acknowledged!!}
%
%ccccccccccccccccccccccccccccccccccccccccccccccccccccccccccccc
\begin{appendix}
%ccccccccccccccccccccccccccccccccccccccccccccccccccccccccccccc
\section{Kerr-resonators with arrays of Josephson junctions}
\label{app:array}
%cccccccccccccccccccccccccccccccccccccccccccccccccccccccccccccccccccc
%
We consider a Kerr resonator, consisting of a linear 1D array of $N$ Josephson junctions (JJ) forming a nonlinear inductance, with first and last junctions capacitively shunted to ground. We will, at first, assume that each of these junctions are sufficiently linear such that we can neglect the nonlinearity of each junctions. Each junction is described by its effective inductance $L_J$ and capacitance $C_J$. Furthermore, we include a parasitic capacitance to ground for each junction, $C_0$. This gives us the (linearized) Lagrangian for the array:
\begin{align}
\mathcal{L}_{array} &= \sum_{n=1}^{N} \frac{C_J}{2} (\dot{\phi}_n - \dot{\phi}_{n+1})^2 + \frac{C_0}{2} \dot{\phi}_n^2 \nonumber\\&\qquad\quad- \frac{1}{2L_{J}} (\phi_n - \phi_{n+1})^2.
\label{eq:Larray}
\end{align}
The terms in Lagrangian corresponding to the end capacitances can be written as,
\begin{align}
\mathcal{L}_{end} = \frac{C_s}{2} \dot{\phi}_1^2 + \frac{C_g}{2} \dot{\phi}_1^2 + \frac{C_e}{2} \dot{\phi}_{N+1}^2,
\end{align}
where $C_s$ is the capacitance to the transmission line which controls the external quality factor of the array resonances, $C_g$ is the shunt capacitance on the first junction, and $C_e$ is the shunt capacitance on the last junction. Including the terms due to  shunt capacitances, the full Lagrangian can then be witten as
\begin{align}
\mathcal{L} &= \mathcal{L}_{array} + \mathcal{L}_{end}\nonumber\\
&= \dot{\vec{\phi}}^{\,T} \frac{\mathbb{C}}{2} \dot{\vec{\phi}} - \vec{\phi}^{\,T} \frac{\mathbb{L}}{2} \vec{\phi}, \label{eq:CLmat}
\end{align}
where we have introduced the symmetric matrices $\mathbb{C}$ and $\mathbb{L}$ along with the vector $\phi = \big\{ \phi_1, \phi_2, \ldots, \phi_{N+1} \big\}^T$

Standing waves across the array now constitute a set of normal modes for the array. Formally, we find these standing mode by diagonalizing the matrix ${\Omega^2 = \mathbb{C}^{-1}\mathbb{L}^{-1}}$, such that the Euler-Lagrange equations for these modes decouple. The eigenvalues of the matrix, $\Omega^2$ are the squares of the normal-mode frequencies of the Kerr resonator. We can similarly define the effective capacitances and inductances for these modes as
\begin{align}
C_k = \vec{v}_k^{\,T} \mathbb{C} \vec{v}_k, && L_k^{-1} = \vec{v}_k^{\,T} \mathbb{L}\vec{v}_k,
\end{align}
where $\vec{v}_k$ are the corresponding eigenvectors of $\Omega^2$.
The eigenfrequencies are now, by construction, given by $\omega_k = 1/\sqrt{L_k C_k}$. Keep in mind that the actual modes are described by the physical phase variable $\phi(t) = \sum_k \phi_k(t) \vec{v}_k$, where $\phi_k(t)$ is a function oscillating with $\omega_k$; thus, the physical amplitude of the phase is in the $\phi_k(t)$ variables and no physical quantify depends on the normalization of $\vec{v}$.
%
%%%%%%%%%%%%%%%%%%%%%%%%%
\subsection{Quantization of the modes}
%%%%%%%%%%%%%%%%%%%%%%%%%
%
Having found the normal modes we can apply a canonical quantization scheme to develop a quantum model for the JJ array~\cite{devoret1995quantum}. This is done by first rewriting the Lagrangian into diagonal form
\begin{align}
\mathcal{L} = \sum_k \frac{C_k}{2} \dot{\phi}_k(t)^2 - \frac{1}{2L_k} \phi_k(t)^2.
\end{align}
This Lagrangian now yields the conjugate variables
\begin{align}
q_k = C_k \dot{\phi}_k^2,
\end{align}
such that we can introduce the quantum operators $\hat{\phi}_k$ and $\hat{q}_k$ satisfying the commutation relation $[\hat{\phi}_k, \hat{q}_k] = i\hbar$. The Hamiltonian for the system is then readily obtained as
\begin{align}
H = \sum_k \frac{\hat{q}_k^2}{2C_k} + \frac{\hat{\phi}_k^2}{2L_k},
\end{align}
which can be recast into the form
\begin{align}
H = \sum_k \omega_k a_k\dag a_k^{\phantom{\dagger}}.
\end{align}
by introducing the ladder operators $(a_k\dag, a_k)$ as
\begin{subequations}
\begin{align}
\hat{\phi}_k  &= \sqrt{\frac{\hbar\omega_k L_k}{2}} (a_k\dag + a_k),\\
\hat{q}_k &= i\sqrt{\frac{\hbar\omega_k C_k}{2}} (a_k\dag - a_k).
\end{align}
\label{eq:phi}
\end{subequations}
The real physical phase variable is now described by the quantum variables, $\hat{\phi} = \sum_k \hat{\phi}_k \vec{\phi}_k$ and we confirm that for a given wave-function of the system the phase-variable is independent of the normalization of $\vec{\phi}_k$.
%
%%%%%%%%%%%%%%%%%%%%%%%%%%%%%%%%%%%%%
\subsection{Reintroduction of the nonlinearity}
%%%%%%%%%%%%%%%%%%%%%%%%%%%%%%%%%%%%%
%
%
%FFFFFFFFFFFFFFFFFFFFFFFFFFFFFFFFFFFFFFFFFFFFFFFFFFFFFFFFFFFFFFFFFFFFFFFFFFFFFFFFFFFFFFFFFFFFFFFFFFFFFFFF

\begin{figure}[t!]
\centering
\includegraphics[width=0.9\columnwidth]{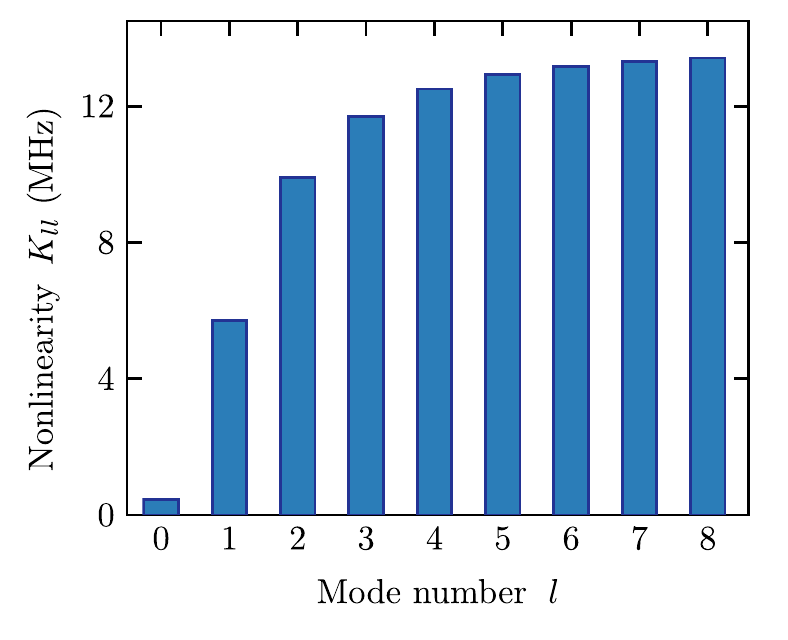}
\caption{Calculated self-Kerr shifts per photon $K_{ll}$ for different modes of the Josephson junction array described in Sec. \ref{sec:exp}.} \label{fig:Kerrshifts}
\end{figure}

%FFFFFFFFFFFFFFFFFFFFFFFFFFFFFFFFFFFFFFFFFFFFFFFFFFFFFFFFFFFFFFFFFFFFFFFFFFFFFFFFFFFFFFFFFFFFFFFFFFFFFFFF
%
The potential energy quadratic in phase variables, as used in Eq. (\ref{eq:Larray}) implicitly assumes small phase excursions $\phi \ll \varphi_0$ (with $\varphi_0 = \hbar/(2e)$) for which Josephson junction potential  $-E_J \cos(\phi / \varphi_0)$ is approximated as $\cos x \approx 1 - x^2$. 
%We may, however, have higher order terms of significance in the system, so now we will reintroduce the nonlinear contributions of the junctions in the array. 
To study the effect of nonlinear contributions of the junctions in the array, we include the next term in the expansion that gives rise to a nonlinear term in the potential, $\mathcal{U}_{nl}$, seen by the phase $\phi$ 
\begin{align}
\mathcal{U}_{nl} = -\frac{1}{24L_J \varphi_0^2} \sum_{n=0}^N \big( \phi_n - \phi_{n+1} \big)^4.
\end{align}
Introducing the variable
\begin{align}
\Delta \phi_k (n) = {v}_k [n] - {v}_k [n+1],
\end{align}
where ${v}_k [n]$ denotes the $n$th entry in the vector $\vec{v}_k$, we can rewrite the potential in terms of the quantum mode operators as,
\begin{align}
\mathcal{U}_{nl} = -\sum_{k_1,k_2,k_3,k_4} \frac{ \prod_{j=1}^4 \hat{\phi}_{k_j}}{24L_J \varphi_0^2} \sum_{n=1}^N \prod_{j=1}^4 \Delta \phi_{k_j}(n). \label{eq:unl}
\end{align}
Generally, such a term will lead to Kerr terms as well as beam-splitter term. The beam-splitter terms are, however, only important between modes with small frequency difference, or if the system is strongly pumped by some external field of appropriate frequency that makes these terms resonant.  Since we are primarily interested in the few photon regime of the lowest modes, which are all well-separated in frequency, we only consider the self-Kerr and the cross-Kerr terms. These can be expressed with the Hamiltonian
\begin{align}
H_{nl} = -\sum_{kl} K_{kl} a_k\dag a_k^{\phantom{\dagger}} a_l\dag a_l^{\phantom{\dagger}}, \label{eq:Hmkerr}
\end{align}
where the Kerr-coefficients are found by inserting Eq.~\eqref{eq:phi} into Eq.~\eqref{eq:unl} and rearranging the terms such that we obtain Eq.~\eqref{eq:Hmkerr} as
\begin{align}
K_{kl} = \frac{2-\delta_{kl}}{4L_J \varphi_0^2} \frac{\hbar \omega_k L_k}{2} \frac{\hbar \omega_l L_l}{2} \sum_{n=1}^N \Delta \phi_k (n)^2 \Delta \phi_l (n)^2, \label{eq:K}
\end{align}
with $\delta_{kl}$ being the Kronecker-delta. Figure \ref{fig:Kerrshifts} shows the Kerr shifts for the first eight array modes, calculated using the parameters for the array presented in Sec. \ref{sec:exp}. We note that the presence of end capacitances loads the array and decreases the mode frequencies significantly \cite{PhysRevLett.109.137002}. As evident from Eq.~(\ref{eq:K}), the mode frequencies and concomitant Kerr shifts per photon can be much larger for an unloaded Josephson array.

As a last important detail, we should mention that the nonlinear coupling between the modes also drags the eigenfrequencies down such that the real mode frequencies become 
\begin{align}
\omega_k' = \omega_k - \sum_l K_{kl}.
\end{align}
This extra contribution is very important as this can shift the frequency of each mode by as much as 2 GHz for parameters similar to the experiment of Sec.~\ref{sec:exp}. The Hamiltonian for the array is, therefore, expressed as
\begin{align}
H = \sum_k \omega_k' a_k\dag a_k^{\phantom{\dagger}} - \sum_{kl} K_{kl} a_k\dag a_k^{\phantom{\dagger}} a_l\dag a_l^{\phantom{\dagger}}.
\end{align}
%

%ccccccccccccccccccccccccccccccccccccccccccccccccccccccccccccc
\section{Classical solution for Kerr oscillator}
\label{app:classical}
%ccccccccccccccccccccccccccccccccccccccccccccccccccccccccccccc
%
We adopt a simple classical description of the driven-dissipative Kerr resonator by replacing the quantum operator $a$ with the classical complex variable $\alpha$ and ignore vacuum fluctuations. This yields the equation of motion (for convenience we change the phase of the drive)
\begin{align}
\dot{\alpha} = -i\Delta \, \alpha + 2iK \, |\alpha|^2 \alpha - \frac{\kappa}{2} \, \alpha + \epsilon. \label{eq:simple_alpha}
\end{align}
In steady state, this leads to the following expression for $n = |\alpha|^2$,
\begin{align}
\varepsilon^2 = \Delta^2 n - 4\Delta K n^2 + 4K^2 n^3 + \frac{\kappa^2}{4} n, \label{eq:clas_ne}
\end{align}
which can be used to find the drive power needed for a given target photon number. We may take the inverse to obtain $n(\varepsilon)$, however, we are not ensured that this function will be single-valued. As a matter of fact, there will always be a set of parameters, $\Delta$ and $K$, for a given $\kappa$ where $n(\varepsilon)$ is multivalued and we can write the highest and lowest $n$'s in this bistable regime as 
\begin{align}
n_{c,\pm} = -\frac{\Delta}{3 K} \Bigg[ 1 \mp \sqrt{1 - \frac{3}{4}\left(1 + \frac{\kappa^2}{4\Delta^{2}}\right)} \;\Bigg]. \label{eq:ncpm}
\end{align}
This expression for $n_{c,\pm}$ can now be inserted into Eq.~\eqref{eq:clas_ne} to get the critical drive power to be in the bistable regime. An alternative approach would be to numerically propagate the dynamical equation, Eq.~\eqref{eq:simple_alpha} until the steady state is reached. In contrast to the multi-valued solution of Eq.~\eqref{eq:clas_ne}, this would only give a single value, but if the parameters are chosen to be in bistable regime the steady state will depend on the initial condition. Experimentally this appears as hysteresis in the system, where, say, a continuous adiabatic change of $\Delta$ will yield different results for increasing or decreasing $\Delta$. 
%
%FFFFFFFFFFFFFFFFFFFFFFFFFFFFFFFFFFFFFFFFFFFFFFFFFFFFFFFFFFFFFFFFFFFFFFFFFFFFFFFFFFFFFFFFFFFFFFFFFFFFFFFF

\begin{figure}[t!]
\includegraphics[width=\linewidth]{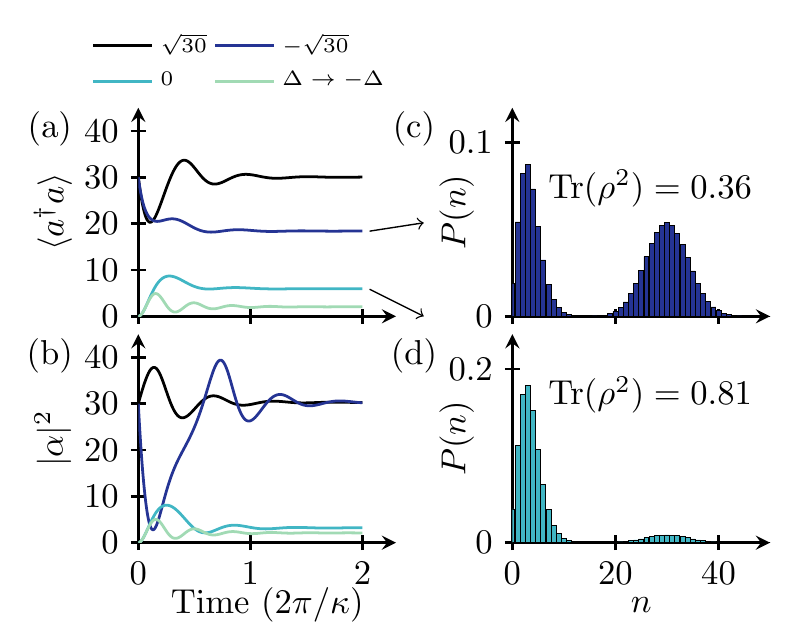}
\caption{The classically bistable regime with $K = \kappa/20$, $\Delta = 2.5\kappa$ and $\epsilon = 4\kappa$. In (a) we see a master equation simulation for an initial coherent state of amplitude $\alpha(0) = \sqrt{30}$ and with a different phase, $\alpha(0) = -\sqrt{30}$. Furthermore we have a simulation starting from vacuum and a simulation from vacuum for two different signs of the detuning. We do the same simulation in (b), but using the classical equation of Eq.~\eqref{eq:simple_alpha}. Panels on the right show the photon number distributions obtained for the final state, where $P(n)$ is the probability to be in the $n$-Fock state, for two different initial states: (c)~$\alpha(0) = -\sqrt{30}$ and (d)~$\alpha(0) =0$ (vacuum).} \label{fig:classical}
\end{figure}

%FFFFFFFFFFFFFFFFFFFFFFFFFFFFFFFFFFFFFFFFFFFFFFFFFFFFFFFFFFFFFFFFFFFFFFFFFFFFFFFFFFFFFFFFFFFFFFFFFFFFFFFF
%
It is worth noting that when $\kappa \gg K$, the dissipation in the resonator dominates the dynamics, destroying all the quantum coherent effects. In this respect, for a Kerr-resonator with $K/\kappa \ll 1$ a classical description of dynamics should suffice. To investigate this further, we compared the results of master equation and classical field equation simulations, shown in Fig.~\ref{fig:classical}. The first thing that we observe in Fig.~\ref{fig:classical} is that for an initial coherent state, $\alpha(0) = \sqrt{30}$, which is very close to the steady state, we have initially a small amount of oscillations in both the quantum [Fig.~\ref{fig:classical} (a)] and classical [Fig.~\ref{fig:classical} (b)] simulations before both converge to the same steady state value for the photon number. Interestingly, the initial phase of the oscillations is very different between the classical and quantum simulations. 

The second observation is that if we change the initial phase of the coherent state to $\alpha(0) = - \sqrt{30}$, the difference between the quantum and the classical simulation becomes pronounced. The quantum simulation shows a small drop in photon number followed by a convergence towards a steady state value different from the one with opposite phase. The classical simulation shows oscillations, since we are far away from the steady state, but eventually the photon number converges to the same steady-state value that we found for the simulation with opposite sign on $\alpha(0)$. Now one might be tempted to interpret this behavior as the mean photon number being very different in the classical case and in the quantum case. Nonetheless, on taking a closer look at the steady state photon number distribution obtained from the quantum simulation [Fig.~\ref{fig:classical}(c)-(d)], we find that this is not the case. We see that the origin of the intermediate photon number comes from a dual-peaked photon number distribution; this can be interpreted as the steady state being a mixed state between the two peaks, each very close to a Poissonian distribution. The classical simulation is only single-valued, so it always selects only one of the solutions, depending on the size of the initial coherent state. We observe a similar behavior, when we start in the vacuum state. Here, there is always a very few number of photons, so the nonlinearity plays a smaller role. The quantum and the classical simulations are very similar, but, due to the quantum fluctuations, there is a small probability for the oscillator to end up in the high photon number state even when it starts in  vacuum [see Fig.~\ref{fig:classical} (d)].

Finally, we see that if we change the sign of the detuning, $\Delta$, the quantum and classical simulation are identical. This is because the sign change moves us away from the bistable regime. So in both cases, the dynamics is deterministic and, since we are always at a low photon number, the nonlinearity again plays a small role. 

\end{appendix}
\bibliography{bt_nourl}
\end{document}